# Prediction of corrosions in Gas and Oil pipelines based on the theory of records

## Mohammad Doostparast
*Iranian Oil Pipeline and Telecommunications Company (IOPTC)*

## Mahdi Doostparast[1]
*Department of Statistics, Ferdowsi University of Mashhad, Mashhad, 91775-1159, Iran*

**Abstract**

Predictions of corrosions in pipelines are valuable. Based on the available data sets, it is critical and useful, for example in preventive maintenance. This paper deals with this problem by two powerful statistical tools, i.e. theory of records and non-homogeneous Poisson process, for modeling location of corrosions. These methods may be used for comparing performances of different pipelines via a quantitative approach. For illustration purposes, we applied the obtained results for a real pipeline in order to prediction the next corrosions based on the available data. Finally, some concluding results and further remarks are given.

# 1. Introduction

Predictions of corrosions positions based on available data is valuable and useful for maintenance in pipeline industries. Estimation of corrosion positions can be used to prevent accidents due to corrosions and also it provides guidelines for managers to allocate properly available resources. For more information, see Meriem-Benziane *et al.* (2017), Ppoola *et al.* (2013), Hu and Cheng (2016), Li *et al.* (2009), Lawless (2002) and references therein. This paper uses the theory of records as well as non-homogeneous Poisson process for modeling the corrosion locations via a statistical approach to capture stochastic behaviors in the phenomenon of corrosions. In this technique, the observed corrosion locations, in a given gas or oil pipeline, have been modeled by the non-homogeneous Poisson process. The estimated parameters may be used for comparison purposes among various pipelines. Also, the models are used to predict the future corrosion positions in similar pipelines. The accuracy of the fitted models is verified by some statistical goodness of fit tests. The obtained results may be used for the assessment in gas and oil pipelines. For illustration purposes, a real data set on an Iranian oil pipeline network is analyzed.

---

[1] The corresponding author;
Emails: doostparast@gmail.com (Mohammad Doostparast); doustparast@um.ac.ir (Mahdi Doostparast).

Therefore, the rest of this paper is organized as follows: The concepts of the non-homogeneous Poisson process (NHPP) and record values are reviewed and explained in Sections 2 and 3, respectively. A power function model for the rate of the NHPP, which is used in pipeline failure prediction, is given in Section 4. The estimate for the above-mentioned rate is stated in Section 5. A real data set on an Iranian oil pipeline is analyzed in Section 6. Finally, Section 7 concludes.

## 2. The non-homogeneous Poisson process

Let $\{N(t), t \geq 0\}$ be the number of occurrences of events (corrosions) in the interval $[0, t]$. Then, $N(t)$ is called a counting process if $N(0) = 0$. Moreover, suppose that

$$P(N(t+h) - N(t) = k) = \frac{e^{-k\lambda(h)} [\lambda(h)]^k}{k!}, \quad k = 0, 1, 2, \cdots \quad t > 0$$

and the increments $N(t_2) - N(t_1)$ and $N(t_4) - N(t_3)$ for all $t_1 < t_2 < t_3 < t_4$ are independent. Then $\{N(t), t \geq 0\}$ is called a Non-Homogenous Poisson Process (NHPP) with the rate function $\lambda(t)$.

### 2.1 Proposition

Let $T_i, \ i = 1, \dots, m$ be the position of the $i$-th corrosion in a pipeline. Then, the joint probability density function (PDF) of $T_1, \dots, T_m$ is: (Arnold et al. 1998)

$$f_{T_1,\dots,T_m}(t_1,\dots,t_m) = \prod_{i=1}^{m} \lambda(t_i) e^{-\int_0^{t_m} \lambda(x)dx} \qquad 0 < t_1 < \cdots < t_m \qquad (1)$$

From Equation (1), one can predict the places of future corrosions. To do this, we need to estimate the rate function λ(t) on the basis of observed $T_1, \dots, T_m$. Here, we use the theory of records and related methods to approximate λ(t). Specifically, let $T_1, \dots, T_m$ denote the first $m$ corrosions and $T_s \ (s > m)$, be s-th future corrosion. One predict for $T_s$ is

$$\hat{T}_s = E(T_s | T_1, \dots, T_m) = \int_{t_m}^{\infty} y f_{(T_s | T_1, \dots, T_m)}(y | t_1, \dots, t_m) dy \qquad (2)$$

where

$$f_{(T_s | T_1, \dots, T_m)}(y | t_1, \dots, t_m) = \frac{[\Lambda(y) - \Lambda(t_m)]^{s-m-1}}{\Gamma(s-m)} \times \frac{f(y)}{R(t_m)} \qquad (3)$$

stands for the conditional PDF of $T_s$ given $T_1, \dots, T_m$. Here,

$$\Lambda(t) = \int_0^t \lambda(x)dx,$$
$$R(t) = exp\{-\Lambda(t)\}$$
$$f(t) = \frac{\partial R(t)}{\partial t}$$
$$\Gamma(a) = \int_0^{\infty} x^{a-1} e^{-x} dx$$

## 3. Theory of records

Let $X_1, X_2, \ldots$ be a sequence of random variables. An observation $X_i$ is called an upper record if $X_i > X_j$ for $i > j$. By definition, $X_1$ is called the first upper records. The joint density function of the first $m$ upper records $R_1, \ldots, R_m$ is:

$$f_{R_1,\ldots,R_m}(r_1, \ldots, r_m) = \prod_{i=1}^{m} \frac{f(r_i)}{1-F(r_i)} e^{-\Lambda(r_m)} \tag{4}$$

provide that the sequence $\{X_i\}_{i\geq 1}$ be independent and identically distributed with the common distribution function $F(x)$ and the density function $f(x)$. For more details, see Dunsmore (1983), Arnold et al. (1998) and references therein. Notice that with $\lambda(t) = f(t)/[1 - F(t)]$ and $\Lambda(t) = \int_0^t \lambda(x)dx$, the distribution of $R_1, \ldots, R_m$ and the place of the first $m$ corrosions following a non-homogeneous Poisson process with the rate function $\lambda(t)$ are identical.

## 4. Power function for λ(t)

In the non-homogeneous Poisson process for corrosion in given pipeline, one should consider a suitable form for the rate function λ(t) according to the effective factors. Also, one can estimate λ(t) by the available data via statistical approaches. To do this, a flexible power form for λ(t) in (1) is assumed in this section. Specifically, let $\lambda(t) = \beta t^{\alpha-1}$ for $t > 0$, $\beta > 0$, $\alpha > 0$. Then λ(t) is increasing (decreasing, constant) for $\alpha > 1$ ($\alpha < 0$, $\alpha = 0$). A graph for λ(t) is given in Figure 1 for some selected value of α and β.

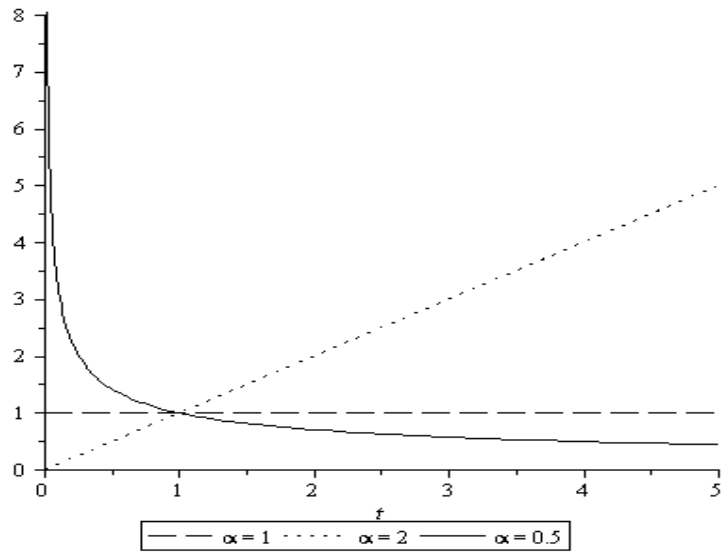

**Figure 1**: The rate function $\lambda(t) = \beta t^{\alpha-1}$ for $\beta = 1$ and $\alpha = 0.5, 1, 2$

Notice that one may use alternative form for λ(t). For more details, see Lawless (2002) and references therein.

# 5. Estimation of the race function λ(t)

As above mentioned, the count process of the corrosion position and upper records are statistically equivalent. Thus, one can use the theory of records to estimate the rate function λ(t) associated to the corrosion location process in a given pipeline and then she can predict the future corrosions by (2). To do this, let $R_1, \ldots, R_m$ be the first *m* corrosion positions. Assuming the power rate function $\lambda(t) = \beta t^\alpha$, for $t \geq 0$, Ahmadi and Doostparast (2006) derived that the maximum likelihood estimates of α and β as

$$\hat{\alpha} = \frac{m}{m \ln(R_m) - \sum_{i=1}^{m} \ln(R_i)} \quad (5)$$

and

$$\hat{\beta} = \frac{m}{R_m^{\hat{\alpha}}} \quad (6)$$

respectively. It is proved that the conditional PDF $R_s$ given $R_1 = r_1, \cdots, R_m = r_m$ is given by (3). By substituting (5) and (6) into (3), one can use the condition expectation of *R* given $R = t_1, \cdots, R_m = t_m$, i.e.

$$\hat{T}_s = \int_{t_m}^{\infty} x f_{(T_s|R_1,\cdots R_m)}(x|t_1, \cdots, t_m, \hat{\alpha}, \hat{\beta})\, dx$$

# 6. Case study
In Table 1, the locations of 18 corrosions in a pipeline (in kilometers) is given.

Table 1: corrosion locations in kilometer

| i | 1 | 2 | 3 | 4 | 5 | 6 | 7 | 8 | 9 |
|---|---|---|---|---|---|---|---|---|---|
| $R_i$ | 0.772 | 2.731 | 3.174 | 16.580 | 25.540 | 25.580 | 32.224 | 32.709 | 34.714 |

| i | 10 | 11 | 12 | 13 | 14 | 15 | 16 | 17 | 18 |
|---|---|---|---|---|---|---|---|---|---|
| $R_i$ | 37.608 | 44.324 | 44.325 | 45.733 | 45.734 | 46.500 | 50.369 | 50.370 | 50.545 |

In this section, we predict 18-th corrosion on the basis of the first observed 17 corrosions. To do this, a NHPP with $\lambda(t) = \beta t^{\alpha-1}$ is assumed. For assessing the appropriate of the NHPP, the random variables $u_i = \log(T_i|T_{i-1})$, $1 \leq i \leq 17$ must follow the exponential distribution (Lawless, 2002). We computed the values of $u_1, \ldots, u_{17}$ for the data set in Table 1 and the P-value of the Kolmogorov-Smirnov goodness of fit test was obtained as 0.564 which supports the NHPP with $\lambda(t) = \beta t^{\alpha-1}$ for the considered data set. The estimates of the parameters are $\hat{\alpha} = 1.1808$ and $\hat{\beta} = 0.1662$. Therefore, the density of $T_{18}$ given $T_1, \ldots, T_{17}$, for $y > 50.545$, is

$$f(y|r_1,\ldots,r_m) = \frac{\left[\hat{\beta}\left(y^{\hat{\alpha}} - r_m^{\hat{\alpha}}\right)\right]^{s-m-1}}{\Gamma(s-m)} \hat{\alpha}\hat{\beta} y^{\hat{\alpha}-1} \exp\left\{-\hat{\beta}\left(y^{\hat{\alpha}} - r_m^{\hat{\alpha}}\right)\right\}$$

$$= \frac{\left[0.1662\left(y^{1.1808} - 102.3154\right)\right]^{s-m-1}}{\Gamma(s-m)} * 0.1962\, y^{0.1808} \exp\left\{-0.1662\left(y^{1.1808} - 102.3154\right)\right\} \quad (7)$$

A point prediction for $T_{18}$ is derived from (2) and (7) with $m = 17$ and $s = 18$ as:

$$\hat{T}_{18} = \int_{r_m}^{\infty} y f(y|r_1,...,r_m) dy$$

$$= \int_{50.370}^{\infty} y \frac{[0.1662(y^{1.1808} - 102.3154)]^{18-17-1}}{\Gamma(18-17)} 0.1962 y^{0.1808} \exp\{-0.1662(y^{1.1808} - 102.3154)\} dy$$

$$= \int_{50.370}^{\infty} 0.1962 y^{1.1808} \exp\{-0.1662(y^{1.1808} - 102.3154)\} dy \tag{8}$$

Another point prediction for $T_{18}$ is the median of the density (7) which is derived by solving the following equation:

$$\int_{r_m}^{\hat{T}_{18}} f(y|r_1,...,r_m) dy = 0.5 \tag{9}$$

After some algebraic manipulations, Equation (9) simplifies as

$$\int_{50.370}^{\hat{T}_{18}} 0.1962 y^{0.1808} \exp\{-0.1662(y^{1.1808} - 102.3154)\} dy = 0.5 \tag{10}$$

By the mathematical package MAPLE version 18, we obtained $\hat{T}_{18} = 52.878$ and $\tilde{T}_{18} = 52.104$, from equations (8) and (9) respectively. A graph of density function $T_{18}$ give $T_1, ..., T_{17}$ is pictured in Figure 2. Any statistical question on $T_{18}$ can be answered by Figure 2. For example, a 95% prediction interval for $T_{18}$ is $[50.4336, 59.4842]$.

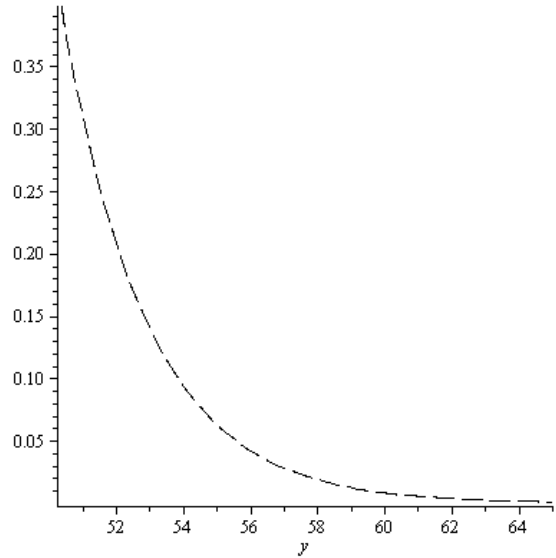

**Figure 2**: The conditional density function 18 given $T_1, \cdots, T_{17}$

| $i$ | 1 | 2 | 3 | 4 | 5 | 6 | 7 | 8 | 9 |
|---|---|---|---|---|---|---|---|---|---|
| Observed corrosions (km) | 0.772 | 2.731 | 3.174 | 16.580 | 25.540 | 25.580 | 32.224 | 32.709 | 34.714 |
| Estimation of α | - | 1.5830 | 1.9180 | 0.6132 | 0.6059 | 0.7264 | 0.7258 | 0.8205 | 0.8801 |

| | | | | | | | | |
|---|---|---|---|---|---|---|---|---|
| Estimation of β | - | 0.4077 | 0.3274 | 0.7150 | 0.702 | 0.5694 | 0.5630 | 0.4573 | 0.3966 |
| Predicted corrosions (km) | - | 3.4901 | 3.6631 | 24.3303 | 35.0046 | 31.7865 | 38.8842 | 37.8170 | 39.1573 |
| | | | | | | | | | |
| $i$ | 10 | 11 | 12 | 13 | 14 | 15 | 16 | 17 | 18 |
| Observed corrosions (km) | 37.608 | 44.324 | 44.325 | 45.733 | 45.734 | 46.500 | 50.369 | 50.370 | 50.545 |
| Estimation of α | 0.9136 | 0.8738 | 0.9532 | 1.0027 | 1.07982 | 1.1366 | 1.1114 | 1.1808 | 1.1808 |
| Estimation of β | 0.3638 | 0.4005 | 0.3233 | 0.2813 | 0.2256 | 0.1910 | 0.2053 | 0.1661 | 0.1662 |
| Predicted corrosions (km) | 41.7606 | 48.9922 | 48.2150 | 49.2410 | 48.7444 | 49.2071 | 53.1849 | 52.8580 | 52.878 |

Table 3: Corrosions (in Kilometers) with the corresponding predictions

## 7. Conclusions and Summary

Predictions of corrosions are valuable in pipeline engineering. These predictions may be used to preventive huge damages as well as to manage the pipeline maintenances. In this paper, the mathematical theory of record values was implemented to model the corrosion locations. The proposed approach applies the theory of nonhomogeneous Poisson process to model the future corrosions on the basis of the observed corrosions in a given pipeline. Also this approach may be used to assess performances of pipelines. For illustration purposes, a real data set on corrosions in an Iranian Pipeline Network was analyzed. This approach provides predictions of corrosion points for pipelines. The magnitudes of the corrosions are also important for pipeline managers. A power function was used as the rate function for corrosion predictions in pipelines. It is worth to deal with complicated forms for the rate functions such as piecewise and polynomial forms. Moreover, estimation of corrosion magnitudes using statistical models is under investigation by the authors. We hope to report the findings in a future paper.